\documentclass{article}

\usepackage{microtype}
\usepackage{graphicx}
\usepackage{subcaption}
\usepackage{booktabs} 
\usepackage{multirow}   
\usepackage[table]{xcolor} 

\usepackage{hyperref}
\usepackage{multirow}

\usepackage[preprint]{icml2026}

\usepackage{amsmath}
\usepackage{amssymb}
\usepackage{mathtools}
\usepackage{amsthm}

\usepackage[capitalize,noabbrev]{cleveref}

\theoremstyle{plain}

\theoremstyle{definition}

\theoremstyle{remark}

\usepackage[textsize=tiny]{todonotes}

\icmltitlerunning{Spatial general purpose audio representation learning}

\begin{document}

\twocolumn[
  \icmltitle{GRAM: Spatial general-purpose audio representations \\
  for real-world environments}



  \icmlsetsymbol{equal}{*}

  \begin{icmlauthorlist}
    \icmlauthor{Goksenin Yuksel}{yyy}
    \icmlauthor{Marcel van Gerven}{yyy}
    \icmlauthor{Kiki van der Heijden}{yyy,sch}

  \end{icmlauthorlist}

  \icmlaffiliation{yyy}{Donders Institute, Radboud University, Nijmegen, The Netherlands}
  \icmlaffiliation{sch}{Mortimer B Zuckerman Institute, Columbia University, New York, United States}

  \icmlcorrespondingauthor{Goksenin Yuksel}{goksenin.yuksel@donders.ru.nl}

  \icmlkeywords{Machine Learning, ICML}

  \vskip 0.3in
]



\printAffiliationsAndNotice{}  

\begin{abstract}

Audio foundation models learn general-purpose audio representations that facilitate a wide range of downstream tasks. While the performance of these models has greatly increased for conventional single-channel, dry audio clips, their success in real-world acoustic environments with reverberation and noise is limited. Furthermore, most audio foundation models ignore the spatial dimension of real-world acoustic environments, ruling out tasks involving sound localization. To address these limitations, we propose GRAM: a general-purpose real-world audio model that employs a multi-channel masked autoencoder to efficiently learn spatial audio representations. We evaluated GRAM and other audio foundation models in a standardized manner on high-quality simulations of naturalistic, spatial acoustic environments as well as recordings of real-world environments and release these two complementary benchmark task suites: NatHEAR and RealSELD. Our results demonstrate that GRAM outperforms all state-of-the-art self-supervised audio foundation models on NatHEAR and the clean, single-channel version HEAR, while using only a fraction of the training data. GRAM also shows state-of-the-art localization performance in simulated environments and generalizes efficiently to real-world recordings in RealSELD.  Taken together, GRAM presents a significant advance toward robust spatial audio foundation models for real-world environments.\footnote{All the code and data is available on \url{https://github.com/labhamlet} and \url{https://huggingface.co/labhamlet}}

\end{abstract}

\section{Introduction}

Despite the complexity and diversity of everyday sound scenes, human listeners effortlessly interact with their acoustic environment in myriad ways. Audio foundation models that perform a similar, human-like range of tasks have received widespread attention~\citep{HEAR,HARES,superb}. While these models demonstrate strong performance on audio benchmarks with minimal fine-tuning (e.g.,~\citep{beats,wav2vec2,mwmae}), they overlook inherent aspects of real-world sound scenes: the spatial dimension, reverberation, and background noise. Specifically, audio foundation models typically lack effectiveness in naturalistic, complex acoustic environments with background noise and reverberation because they are primarily trained on dry, large-scale sound datasets such as AudioSet~\citep{audioset} and Librispeech~\citep{librispeech}.

Crucially, the lack of spatial information in audio embeddings precludes sound localization and the use of spatial sound features to improve performance on complex listening tasks, such as audio scene analysis. Audio scene analysis refers to the separation of overlapping sound waves in complex multi-source sound scenes and the subsequent grouping of the frequency components into distinct auditory objects~\citep{bregman1984auditory,bizley2013and}. In humans, such audio scene analysis is aided by spatial cues~\citep {bizley2013and,heijden2019cortical}. And, incorporating spatial knowledge into universal audio embedding models is also expected to benefit downstream tasks that require ambient intelligence and acoustic awareness, such as acoustic scene understanding.


To address these limitations of audio foundation models for real-world environments, we present GRAM (General-purpose, Real-world Audio Model). GRAM is a self-supervised, multi-channel masked auto-encoder model that efficiently learns spatial general-purpose audio representations from multi-channel audio clips. To train GRAM, we developed a custom pipeline that uses the Soundspace 2.0 platform~\citep{chen22soundspaces2} to simulate high-quality real-world sound scenes. Further, to promote the systematic evaluation of audio foundation models on naturalistic sound scenes, we introduce two complementary benchmark suites: NatHEAR and RealSELD.

NatHEAR is an extension of the HEAR benchmark suite that includes simulated, real-world versions of the downstream tasks. Additionally, it includes two sound localization tasks and reverberation time (T60) estimation tasks. NatHEAR plays a crucial role in assessing the robustness of audio foundation models to controlled noisy and reverberant conditions. RealSELD comprises real-life datasets for sound event localization and detection (SELD) tasks collected from previous DCASE challenges~\citep{dcase}. We embed these evaluation tasks into the standardized HEAR setup, therefore unifying the evaluation of SELD performance of audio foundation models in challenging real-world scenarios. 


Empirical results demonstrate that GRAM efficiently learns robust, general-purpose spatial audio representations, outperforming all state-of-the-art audio foundation models and speech models on HEAR and NatHEAR. GRAM excels at complex tasks, such as audio scene analysis, and achieves strong sound localization performance, outperforming even supervised models trained with auxiliary spatial features. Finally, GRAM demonstrates robust transfer to recordings of real-world sound scenes, as evidenced by RealSELD performance, thereby overcoming the need for extensive domain-specific adaptations. Taken together, our key contributions can be summarized as:  

\textbf{General-Purpose Audio Foundation Model (GRAM):} We present GRAM, a multi-channel masked auto-encoder that shows state-of-the-art performance on a human-like range of tasks in naturalistic sound scenes, including sound localization. GRAM is the first audio foundation model available in both binaural and four-channel Ambisonics audio formats.

\textbf{A large-scale dataset for high-quality simulations of real-world sound scenes}: We release the complete set of binaural room impulse responses (BRIRs) and ambisonics room impulse responses (ARIRs) corresponding to 85,000 naturalistic sound scenes that we used for our naturalistic training pipeline.

\textbf{NatHEAR and RealSELD:}  To enable the systematic evaluation of audio foundation models in complex acoustic environments, we introduce the NatHEAR and RealSELD. NatHEAR extends the HEAR framework with simulated naturalistic scenes and novel spatial tasks, while RealSELD establishes the first benchmark for assessing pre-trained audio embeddings on real-world SELD tasks.

\section{Related Work}
\begin{figure*}[!htb]
    \centering
    \includegraphics[width=\textwidth]{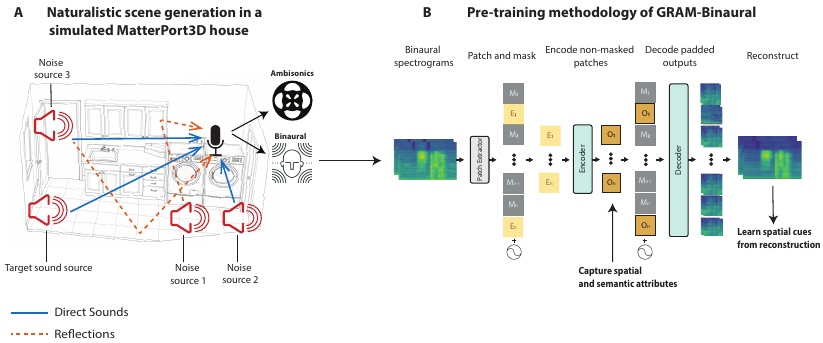}
    \caption{\textbf{Proposed self-supervised approach for training GRAMs on naturalistic binaural scenes}. (A) We generate binaural and ambisonics naturalistic scenes using the SoundSpace2.0 simulator ~\citep{chen22soundspaces2} in MatterPort3D houses. (B) MAE approach for learning audio representations with spatial attributes. For the ambisonics spectrograms, the methodology remains the same, except that the inputs now include 4-channel mel-spectrograms and intensity vectors (IVs). }
    \label{fig:approach}
\end{figure*}
\textbf{Supervised audio representation learning:}
Supervised methods for audio representation learning have achieved notable success in recent years. Approaches such as AST~\citep{ast}, PaSST~\citep{koutini22passt} and HTS-AT~\citep{hts} have Transformer-based architectures as a backbone, for example ViT~\citep{dosovitskiy2020vit} and Swin Transformer~\citep{liu2021Swin}. To mitigate the need for large annotated datasets, some of these approaches rely on pre-trained image models (e.g., PSLA~\citep{psla}). Question-and-answer models constitute a more recent category of supervised approaches that integrate audio representation learning with large language models (for example, Spatial-AST~\citep{spatialast} and Qwen-Audio~\citep{chu2023qwen})res large-scale annotated datasets and is sub-optimal for learning general-purpose audio representations that generalize across tasks. 

\textbf{Self-supervised audio representation learning:}
Self-supervised audio representation learning approaches aim to learn robust audio representations that generalize across a wide range of tasks. Masking-based approaches utilizing transformer backbones to reconstruct masked patches of input spectrograms currently constitute the predominant approach, including (SSAST~\citep{gong2022ssast}), MSM-MAE~\citep{masked-modelling}, MaskSpec~\citep{MaskSpec}, MAE-AST~\citep{mae-ast}, and Audio-MAE~\citep{audiomae} achieves state-of-the-art performance on the HEAR benchmark by using multi-window local-global attention in the decoder. Recently, SSAM~\citep{audiomamba} utilized a Mamba~\citep{mamba} architecture in their encoder and achieved similar performance as MWMAE. In contrast to masked auto-encoders, BEATS~\citep{beats} employs a masking-based approach based on latent embeddings extracted by an acoustic tokenizer. Finally, successful self-supervised approaches that do not rely on masking at all include contrastive learning frameworks such as COLA~\citep{COLA}. 

Another category of self-supervised audio representation models focuses specifically on speech representations, using generative, predictive, or contrastive learning~\citep{mohamed2022self}. These speech models are typically trained on datasets such as Librispeech~\citep{librispeech} or LibriLight~\citep{librilight} and include state-of-the-art models such as Wav2Vec2~\citep{wav2vec2}, HuBERT~\citep{hubert}, and WavLM~\citep{wavlm}. However, while these models excel at speech-based tasks, they do not necessarily generalize well to non-speech sounds and non-speech tasks~\citep{HEAR}. Crucially, none of the existing self-supervised approaches for audio or speech representation learning optimize for performance in real-world sound scenes that are spatial, reverberant, and noisy.
\label{related-work}

\section{Materials and Methods}
\label{methdology}

\textbf{Simulating real-world acoustic scenes:} A room impulse response (RIR) captures room-specific acoustic properties such as reverberation. We utilized high-resolution, detailed 3D meshes of houses with various architectural characteristics from Matterport3D~\citep{Matterport3D} to simulate RIRs for many different rooms in each house with the Monte Carlo ray tracing RIR simulator provided by SoundSpaces 2.0~\cite{chen22soundspaces2}. SoundSpaces 2.0 combines simulated RIRs with a head-related transfer function (HRTF)~\citep{CIPIC} to generate a binaural RIR (BRIR) or, with an ambisonics microphone configuration, an ambisonics RIR (ARIR). BRIRs capture both room acoustic properties and human spatial hearing characteristics introduced by the shape of the ears, head and torso, while ARIRs capture room acoustic properties as well as the spatial cues encoded in first-order Ambisonics.

\textbf{Components of simulated real-world scenes:} Matterport3D contains scans of 90 houses. We discarded five houses for which the meshes were not of sufficient quality. For each of the remaining 85 houses, we simulated 1,000 real-world scenes. Each scene comprised a randomly sampled listener location, a sound source location, and a noise source location within the room. For BRIRs (binaural), we randomly sampled head orientation from a range [0\textdegree, 360\textdegree]). We placed the sound source at a randomly sampled position relative to the listener or microphone (distance range [1.5 m, 5 m]; azimuth range [0\textdegree, 360\textdegree]; elevation range [-90\textdegree, +90\textdegree]). Noise was either localized (50\% of the scenes) or diffuse (50\% of the scenes). For localized noise, we randomly sampled a single location within the room. For diffuse noise, we randomly sampled three, four, or five locations in the room. We then rendered a set of RIRs to describe all components in the naturalistic scene. Given sound source location $s$, listener (microphone) location $r$, and receiver head orientation $\theta$, we rendered RIRs describing the sound path from the source to the listener (microphone) as $\text{BRIR}(s, r, \theta)$ and as $\text{ARIR}(s, r, \theta)$.  Given a number of noise sources $n_{i}$ at noise source location $\phi_i$, listener location $r$, and receiver head orientation $\theta$, we rendered the RIR describing the path from the noise source(s) to the listener as $\text{BRIR}_{i}(\phi_i, r, \theta)$ and as $\text{ARIR}(\phi_i, r, \theta)$. This procedure resulted in a total of 85,000 sets of BRIRs as well as 85,000 sets of ARIRs (see \autoref{appendix: brir-spec} for all parameters).

\subsection{GRAM framework}

GRAM learns spatial audio representations by reconstructing masked multi-channel binaural and ambisonics spectrogram patches. Importantly, GRAM reconstructs crucial localization cues, such as interaural level differences (ILDs) for binaural scenes and intersity vectors (IVs) for ambisonics scenes, thereby learning to encode the necessary spatial information. We refer to the binaural version as GRAM-Binaural, and the ambisonics version as GRAM-Ambisonics. First, a patch extractor consisting of a single convolutional layer with convolutional filters divides each multi-channel spectrogram into $n$ non-overlapping patches $P_1, \ldots, P_n$ with $P_i \in \mathbb{R}^{C \times T \times F}$, and embeds each patch into a linear patch embedding $E_i \in \mathbb{R}^{768}$ (\autoref{fig:approach}). Non-masked patch embeddings are input to the encoder, for which we selected the 12-layer ViT-Base (ViT-B) Transformer~\citep{dosovitskiy2020vit} similar to \citet{audiomae, mwmae}. The encoder outputs patch representations $O_i \in \mathbb{R}^{768}$ for $i = 1, \ldots, n$, where $n$ is the number of unmasked patches. Finally, a Transformer decoder with local-global attention~\citep{mwmae} followed by a linear head takes all patch representations $O_1, \ldots, O_n$ as well as all masked patches $M_1, \ldots, M_n$ to reconstruct the multi-channel spectrogram from last layer embeddings.

\subsection{Pre-training}

\begin{figure*}[!t]
    \centering
    \includegraphics[width=\textwidth]{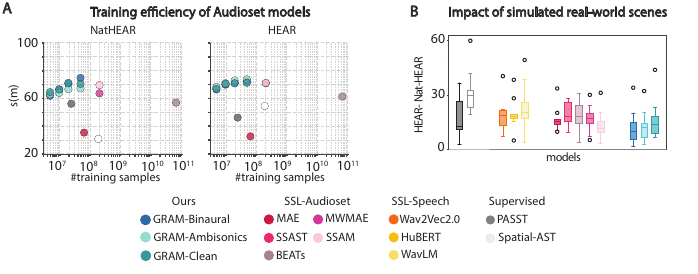}
    \caption{\textbf{Downstream model performance}. (A) NatHEAR and HEAR performance as a function of training data. (B) Difference in performance on HEAR and NatHEAR (excluding the DCASE-2016 task). Box limits reflect first and third quartile, center line the median.}
\label{fig:nathear}
\end{figure*}

\textbf{Online mixing of naturalistic sound scenes:}
\label{synth}
The 85,000 naturalistic scenes were split into a train set of 70,000 scenes (corresponding to 70 Matterport3D houses), and a test set of 15,000 scenes (15 Matterport3D houses) for downstream evaluation (see Section~\ref{downstream_exp}). We used the 70,000 naturalistic scenes in the train set to generate naturalistic scenes for all audio clips in the unbalanced AudioSet training set (10-second soundtracks from 1.74 million YouTube videos~\citep{audioset}). Specifically, during training, we randomly paired an AudioSet clip with a noise sound clip from the WHAMR! background noise database~\citep{Maciejewski2020WHAMR}. WHAMR! Noise clips longer than 10 s were trimmed to 10 s duration, and a linear fade-in/fade-out of 200 ms was applied to every noise clip before mixing of the sound scene. 

To create a naturalistic sound scene, we then convolved the AudioSet clip with either BRIR$(s,r,\theta)$ for GRAM-Binaural or ARIR$(s,r,\theta)$ for GRAM-Ambisonics to obtain $T$. Similarly, we convolved the WHAMR! noise clip with the BRIR$(\phi_i, r, \theta)$ to obtain $N_{i}$. In naturalistic scenes with diffuse background noise, the diffuse noise field $D$ was generated by summing the noise clips $D = \sum_{i=1}^{M} N_i$ where $N_i$ are individual noise clips and $M$ is the total number of noise clips. The naturalistic sound scene $S$ was then computed as $S = T + bN$ for scenes with localized noise and as $S = T + bD$ for scenes with a diffuse noise field. Here, $b$ is a scaling parameter that mixes target and noise clips at a signal-to-noise ratio (SNR) ranging from +5 dB to +40 dB.

\textbf{Input features:}
We transformed the channels of each sound scene into log-scale mel spectrograms using 128 mel filters in the frequency range of 50-16000 Hz with a 25 ms Hanning window and 10 ms hop length, resulting in spectrograms of dimension $1001 \times 128$. We added zero padding to achieve a $1024 \times 128$ dimension.  For GRAM-Ambisonics, following \citet{elsa, ambisonic_IVs}, we  extracted intensity vectors utilizing the equation below:
 \begin{equation}
     \label{eq:iv}
     I_\mathrm{active}(t,f) = \Re \left[ A_\text{0,0}^{*}(t,f) \begin{pmatrix}A_\text{1,-1}(t,f)\\ A_\text{1,0}(t,f)\\ A_\text{1,1}(t,f) \end{pmatrix} \right]
 \end{equation}
 
 Where $A_{n,m}$ are the $n^\textrm{th}$ and $m^\textrm{th}$ order and mode of the ambisonics signal corresponding to its omnidirectional ($W$) and three dipole $(Z,Y,X)$ components,  and $(\cdot)^*$ denotes complex conjugation. IVs are scaled to unit norm. 
We concatenated mel spectrograms and intensity vectors, resulting in input $x = [x_\text{mel}, IVs]$. 

\textbf{In-batch sampling:} As the online mixing of naturalistic acoustic scenes is computationally expensive due to multiple long convolutions, we used a random in-batch sampling procedure to increase the effective batch size in a computationally efficient manner. We randomly sampled 16 partially overlapping segments of 2 seconds to create 16 samples of dimension 200 $\times$ 128. This increases the original batch size of 96 to an effective batch size of 1536.

\textbf{Patch extraction and masking:}
\label{masking_explanationss}
For pre-training, we divided the binaural spectrogram into $P_i \in \mathbb{R}^{2 \times 8 \times 16}$,  and ambisonics spectrograms into $P_i \in \mathbb{R}^{7 \times 8 \times 16}$ patches. We used an adapted version of the mask-based framework of MWMAE~\citep{mwmae}, randomly selecting a subset of $n$ patches $M_1, \ldots, M_n$ for $i = 1, \ldots, n $ for masking (masking ratio = 0.8) and replacing their embedding with a learnable mask token. Finally, we added fixed sinusoidal positional embeddings to all embedded patches.

\textbf{Decoder with local-global attention: } The decoder takes as input both the unmasked patches $O_1, \ldots, O_n$ with $O_i \in \mathbb{R}^{768}$, and the masked patches $M_1, \ldots, M_n$ with $M_i \in \mathbb{R}^{768}$ as well as fixed sinusoidal positional embeddings for each patch (\autoref{fig:approach}). To implement local-global attention~\citep{mwmae}, we selected window sizes of [2, 5, 10, 25, 50, 100, 0, 0]. Here, 0 signifies global plain attention.

\subsection{Ablations} 
To identify key factors for successful learning of spatial general-purpose audio representations, we conducted a series of ablation experiments. For GRAM-Binaural, we swapped the Transformer encoder with a Mamba encoder~\citep{mamba, audiomamba}. To ensure that computational overhead and model capacity were comparable between the Transformer and Mamba encoders, we used similar parameter counts. We also tested the impact of mask type for GRAM-Binaural, comparing patch-based masking and time-based masking. For time-based masking, patches were defined as $P_i \in \mathbb{R}^{2 \times 2 \times 128}$, spanning the entire frequency range. For time-based masking, we used window sizes [2, 5, 10, 25, 50, 0, 0, 0] to implement local-global attention in the decoder. Furthermore, for both GRAM-Binaural and GRAM-Ambisonics, we assessed the optimal ratio ($\lambda$) between simulated real-world sound scenes and clean, dry sound clips in pretraining data for $\lambda = 0.0, 0.25, 0.5, 0.75, 1.0$.  Finally, we examined various masking ratios [0.4, 0.6, 0.8, 0.9] and in-batch sampling factors [4, 8, 16] for GRAM-Binaural. For all ablations, GRAM-Binaural and, if applicable, GRAM-Ambisonics were trained with the same parameters specified above, except for the masking-ratio ablation, where we reduced the effective batch size from 1536 to 384 to alleviate out-of-memory errors.

\section{Downstream evaluation}
\label{downstream_exp} 


\textbf{HEAR}: This benchmark task suite ~\citep{HEAR} includes a wide range of tasks to evaluate the downstream performance of audio representation models~\citep{HEAR}. We selected the same subset of HEAR tasks as previously used in~\citep{mwmae}, but added HEAR's time-stamp-based sound event detection task DCASE-2016 Task 2~\citep{Mesaros2018_TASLP} to enable in-depth evaluation of audio scene analysis capabilities. 

\textbf{NatHEAR - Simulated naturalistic sound scenes}: To test performance in a wide range of noisy, spatial, naturalistic scenes, we generated NatHEAR. This benchmark task suite contains the same tasks as the original HEAR, but the sound clips are converted to high-quality simulations of naturalistic scenes rather than clean, single-channel scenes. NatHEAR exists in two audio formats: a two-channel binaural format and a four-channel first-order Ambisonics format. We furthermore included sound localization and T60 estimation tasks in NatHEAR across two audio domains. Specifically, we used the SC-5 as an example of the speech domain and ESC-50 as an example of the environmental sound domain. These localization tasks are modeled as a multi-output regression task in which model outputs represent the estimated 3D Cartesian coordinates $ [x, y, z] $ on the unit sphere~\citep{adavanne2018sound}. Further, T60 tasks are modeled as regression tasks with continuous outputs.

\textbf{RealSELD - recordings of real-world sound scenes}: To evaluate performance in real-world scenes in a standardized manner, we created RealSELD. This benchmark task suite comprises real-life datasets for sound event localization
and detection (SELD) tasks collected from previous
DCASE challenges~\citep{dcase} that include both static and moving sound sources, varying in level of complexity. The datasets included are TUT-2018~\citep{seldnet}, TAU-2019~\citep{tau2019}, TAU-NIGENS-2020~\citep{tau2021}, TAU-NIGENS-2021~\citep{tau2021} and STARSS23~\citep{STARSS23}. More information regarding the datasets can be found in Appendix \autoref{tab:RealSELD-datasets}. By embedding these
evaluation tasks into the standardized HEAR setup, we present a standardized evaluation of SELD performance for audio foundation models in challenging real-world scenarios.

\subsection{Evaluation methodology}
\textbf{Downstream task evaluation:}  Following the HEAR protocol~\citep{HEAR} for downstream task evaluation, we extracted embeddings from the frozen pretrained models and subsequently trained a shallow downstream classifier on these embeddings to assess the extent to which the learned representations generalize across a broad range of tasks. We applied this set-up to all benchmark suites - that is, HEAR, NatHEAR and RealSELD - to create a uniform evaluation set-up (for an overview of all tasks, see \autoref{hear_appendix_tasks}). The procedure for the GRAM embedding extraction is described in \autoref{gram_embeddings}. 

To test downstream SELD performance on the real-life recordings in benchmark task suite RealSELD (resampled to 32 kHz), we formulated the localization and detection task following the commonly used Activity-Coupled Cartesian Direction of Arrival (ACCDOA) framework~\citep{shimada2021accdoa}. This framework jointly models sound event detection and localization across target sound classes. Further, to adapt the RealSELD datasets to the HEAR pipeline, we processed the ground truth timestamp labels into fixed-duration segments. For static sources, we extracted start and end times alongside fixed azimuth and elevation coordinates. For moving sources, we extracted active segments and their corresponding time-varying directions. In frames where no sources are active, the ACCDOA target is defined as a zero vector. Crucially, the datasets in RealSELD utilize 100 ms segments while GRAM operates on 80 ms segments (\autoref{gram_embeddings}). To temporally align GRAM with the RealSELD labels, we applied average pooling to the model representations. We then extract the time-stamp embeddings and their corresponding ground truth labels using the HEAR evaluation kit, and linear probe the extracted information using the HEAR evaluation protocol.

\textbf{Impact of real-life recordings during training:} In addition to the evaluation of the direct transfer of GRAM-Ambisonics to real-life recordings described above, we tested the impact of fine-tuning GRAM-Ambisonics on real-life recordings, as well as training GRAM-Ambisonics from scratch on real-world recordings. To this end, we used the most challenging task in RealSELD, namely the STARSS23 dataset (an extended version with additional training data,similar to~\citet{STARSS22}). To fine-tune GRAM-Ambisonics, we initialized the model using pre-trained weights. Both fine-tuning and training from scratch were implemented with a batch size of 512 and 100 training epochs. All other experimental settings followed the SELD baseline model~\citep{STARSS22}. 

\textbf{Matching audio formats:} The input audio format  varies across audio foundation models and benchmark task suites. Hence, to adapt the audio format of the multi-channel task suites NatHEAR and RealSELD to the single-channel models, we selected the omnidirectional channel $W$ of the first-order Ambisonics~\citep{ambisonics} as their model input. Further, to adapt the audio format of single-channel task suite HEAR to the multi-channel models GRAM-Binaural and GRAM-Ambisonics, we duplicated the original HEAR single-channel spectrograms to generate two- or four-channel audio inputs.
\begin{table*}
\renewcommand{\arraystretch}{1.2}
\setlength{\tabcolsep}{3pt} \caption{Performance on NatHEAR. Reported values reflect the average performance $\pm$ standard deviation, calculated using $n$-fold cross-validation as specified by the HEAR. Bold numbers indicate the best
performing model on the specific task. Grayed-out rows indicate supervised models. Tasks are specified in \autoref{hear_appendix_tasks}. \textbf{LS}: LibriSpeech 960 h, \textbf{AS}: AudioSet}
\label{tab:naturalistic_audio_benchmarks}
\centering
\resizebox{\textwidth}{!}{
\begin{tabular}{ll||cccc||ccc||cccc||cc}
\toprule
& & \multicolumn{4}{c||}{\textbf{Acoustic Events and Scene Analysis}}  & \multicolumn{3}{c||}{\textbf{Speech}} & \multicolumn{4}{c||}{\textbf{Music}} & \multicolumn{1}{c}{}\\
\textbf{Model} & \textbf{Corpus} & \textbf{DCASE} & \textbf{FSD50K} & \textbf{LC} &  \textbf{ESC-50} & \textbf{CD} & \textbf{VL} & \textbf{SC-5}   & \textbf{NS} &  \textbf{BO} & \textbf{Mri-S} & \textbf{Mri-T} & \textbf{s(m)} & \textbf{Avg.} \\
\hline

HEAR-Naive & - & 26.5 & 8.7 & $ 27.4 \pm 1.6 $  & $ 17.2 \pm 2.2 $  & $ 32.3 \pm 2.2 $  & $ 11.7 \pm 2.2 $  & 12.0 & 75.6 & $ 84.3 \pm 4.5 $ & $ 68.6 \pm 1.3 $  & $ 60.5 \pm 1.3 $  & 0.0 & 38.6\\
\hline
Wav2Vec2 & LS & 32.0 &  23.0 & $ 54.6 \pm 1.9 $ & $ 36.4 \pm 2.9 $ & $ 48.6 \pm 0.6 $ & $ 27.2 \pm 1.6 $ &  78.9 &  15.2 & $ 71.2 \pm 6.4 $ & $ 75.7 \pm 0.5 $ & $ 45.9 \pm 0.6 $ & 32.5 & 46.2 \\
HuBERT & LS & 57.6 &  26.6 & $ 52.5 \pm 2.2 $ & $ 49.5 \pm 2.2 $ & $ 57.4 \pm 1.1 $ & $ 46.8 \pm 3.4 $ &  89.2 &  16.0 & $ 77.1 \pm 6.0 $ & $ 78.2 \pm 0.7 $ & $ 52.4 \pm 1.6 $ & 45.2 & 54.8 \\
WavLM & LS &  25.3 &  20.5 & $ 52.1 \pm 0.6 $ & $ 41.4 \pm 2.1 $ & $ 52.3 \pm 1.5 $ & $ \textbf{ 47.9} \pm 4.6 $ &  89.9 &  11.2 & $ 61.4 \pm 7.2 $ & $ 69.3 \pm 0.9 $ & $ 39.0 \pm 2.0 $ & 37.8 & 46.4 \\
\hline
MAE & AS & --  & 27.9 & $ 53.2 \pm 1.0 $ & $ 65.7 \pm 1.2 $ & $ 48.5 \pm 1.3 $ & $ 19.0 \pm 1.5 $ & 57.4 & 53.4 & $ 79.2 \pm 7.8 $ & $ 81.0 \pm 4.9 $ & $ 56.5 \pm 12.3 $ & 34.5 & 54.2 \\
SSAST & AS + LS & --  & 15.6 & $ 41.6 \pm 2.4 $ & $ 44.8 \pm 1.0 $ & $ 39.7 \pm 2.9 $ & $ 12.7 \pm 1.3 $ & 19.9 & 52.0 & $ 81.8 \pm 3.6 $ & $ 76.5 \pm 3.6 $ & $ 64.6 \pm 1.5 $ & 17.5 & 44.9 \\
BEATs & AS & --  & 46.5 & $ 63.7 \pm 1.2 $ & $ 72.6 \pm 3.9 $ & $ 54.8 \pm 1.6 $ & $ 27.5 \pm 4.3 $ & 83.5 & 54.2 & $ 70.3 \pm 6.2 $ & $ 83.2 \pm 1.0 $ & $ 71.0 \pm 1.4 $ & 55.7 & 62.7 \\
MWMAE & AS & 83.8 & 44.3 & $ 64.8 \pm 1.1 $ & $ 69.7 \pm 5.6 $ & $ 59.3 \pm 1.0 $ & $ 31.8 \pm 1.8 $ & 86.7 & 59.2 & $ 77.1 \pm 3.6 $ & $ 90.1 \pm 0.8 $ & $ 73.9 \pm 0.6 $ & 62.5 & 67.3 \\
SSAM & AS & 70.0 & 46.0 & $ 63.2 \pm 1.1 $ & $ 73.1 \pm 2.4 $ & $ 62.3\pm 1.0 $ & $ 38.8 \pm 2.6 $ & 86.2 & 65.4 & $ 84.3 \pm 7.0 $ & $ 92.6 \pm 0.4 $ & $ 76.8 \pm 1.0 $ & 68.4 & 68.9\\
\rowcolor[gray]{0.9} GRAM-Binaural & AS & \textbf{93.0} & \textbf{52.8} &  \textbf{72.3} $\pm 0.7 $   & $\textbf{82.6} \pm 3.2 $   & $ \textbf{63.3} \pm 1.3 $ & $ 35.1 \pm 3.8 $   & \textbf{91.0} & \textbf{67.6} & $ \textbf{85.6} \pm 5.1 $   & $ \textbf{91.7} \pm 0.9 $   & $ \textbf{78.3} \pm 1.3 $   & \textbf{74.8} & \textbf{73.9}\\
\rowcolor[gray]{0.9} GRAM-Ambisonics & AS & 90.2 & 49.5 & $ 68.8 \pm 0.9 $   & $ 79.4 \pm 2.7 $   & $ 61.4 \pm 0.9 $ & $ 36.4 \pm 4.2 $   & 87.2 & 64.6 & $ 83.4 \pm 4.7 $   & $ 91.3 \pm 0.6 $   & $ 78.1 \pm 1.4 $   & 70.5 & 71.8 \\
\rowcolor[gray]{0.9} GRAM-Clean & AS & 90.9 & 50.5 & $ 66.4 \pm 0.8 $ & $ 80.0 \pm 2.4 $ & $ 62.0 \pm 1.3 $ & $ 32.2 \pm 2.3 $ & 87.3 & 65.2 & $ 82.2 \pm 5.6 $ & $ 90.2 \pm 0.8 $ & $ 75.1 \pm 0.7 $ & 67.3& 71.1\\
\hline
\color{gray}PASST & \color{gray}AS & \color{gray}--  & \color{gray} 56.9 & \color{gray}$ 52.1 \pm 1.9 $ & \color{gray}$ 89.7 \pm 2.1 $ & \color{gray}$ 49.9 \pm 1.0 $ & \color{gray}$ 18.4 \pm 2.3 $ & \color{gray}61.1 & \color{gray}16.0 & \color{gray}$ 93.6 \pm 4.0 $ & \color{gray}$ 85.5 \pm 1.7 $ & \color{gray}$ 55.6 \pm 3.0 $ & \color{gray}56.2 & \color{gray}57.9 \\
\color{gray}Spatial-AST & \color{gray} AS & \color{gray}--  & \color{gray}40.0 & \color{gray}$ 49.9 \pm 1.5 $ & \color{gray}$ 70.1 \pm 3.3 $ & \color{gray}$ 41.6 \pm 0.5 $ & \color{gray}$ 11.7 \pm 2.7 $ & \color{gray}54.8 & \color{gray}50.2 & \color{gray}$ 77.1 \pm 2.8 $ & \color{gray}$ 77.7 \pm 0.9 $ & \color{gray}$ 55.0 \pm 1.6 $ & \color{gray}30.9 & \color{gray}52.8\\
\bottomrule
\end{tabular}
}
\end{table*}

\begin{figure*}[!t]
    \centering
    \includegraphics[width=\textwidth]{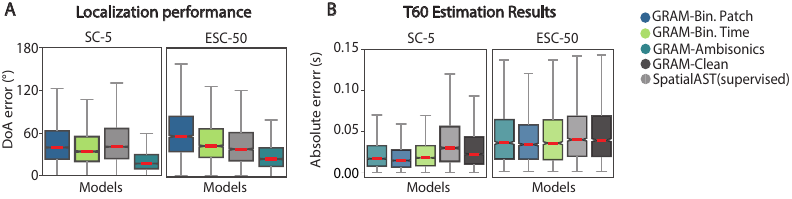}
    \caption{\textbf{Sound localization and T60 estimation in simulated real-world sound scenes}. (A) Direction of arrival (DoA) error.  (B) T60 estimation absolute error. Boxes indicate first and third quartile; center line: median; whiskers: 1.5 times the interquartile range.}
    \label{fig:local}
\end{figure*}

\subsection{Performance metrics}

\textbf{HEAR and NatHEAR}: For each model $m$, we use two metrics to assess overall performance across for each of these task suites: The average performance across all tasks and the metric $s(m)$, which reflects a model's improvement relative to the maximum improvement over the baseline achieved by the current state-of-the-art model, averaged across all tasks in the benchmark. This metric effectively ranks model improvements relative to the baseline as a function of the current maximum improvement (see \autoref{appendix:genmetric}). We use the HEAR-Naive baseline based on mel-spectrograms as baseline here ~\citep{HEAR}. 

\textbf{Evaluation of SELD tasks on real-life recordings:} For dynamic scenes, we used four joint localization and detection metrics for the RealSELD dataset~\citep{eval_overview}, which are widely-used in audio-only SELD tasks~\citep{STARSS22}. These include two metrics for location-aware detection (error rate ER$_{20^{\circ}}$ and F-score F$_{20^{\circ}}$), and two metrics for class-aware localization (localization error LE$_{CD}$ and localization recall LR$_{CD}$). For static scenes, inline with the reported baseline results~\citep{adavanne2018sound}, we used the localization independent version of these metrics. See \autoref{appendix:accdoa} for details on evaluation metrics and the ACCDOA training framework.


\section{Results}

\textbf{Downstream tasks in NatHEAR and HEAR: }\autoref{tab:naturalistic_audio_benchmarks} demonstrates that GRAM-Binaural and GRAM-Ambisonics learn robust general-purpose audio representations, outperforming all other self-supervised audio representation models on NatHEAR. Moreover, (\autoref{fig:nathear} A shows that GRAM requires substantially less training data to achieve this performance level). Further, all GRAMs surpass all other self-supervised audio representation models on the dry, non-spatial and clean sound scenes in HEAR (\autoref{tab:dry_audio_benchmarks}). GRAM-Clean achieved state-of-the art performance, followed by GRAM-Binaural and GRAM-Ambisonics. Comparing the performance of the GRAM models on NatHEAR and HEAR highlights two key findings: (1) The success of GRAM-Binaural and GRAM-Ambisonics on NatHEAR demonstrates the impact of the naturalistic training pipeline for downstream task performance in real-world scenes (\autoref{fig:nathear}B); and (2) The superior performance of GRAM-Binaural and GRAM-Ambisonics on HEAR relative to other audio representation models indicates that the naturalistic training pipeline does not degrade downstream performance on clean, dry sound scenes.

\textbf{Sound localization and T60 estimation in NatHEAR:} 
\label{locresults} \autoref{fig:local}A shows that GRAMs exhibit excellent localization performance in simulated real-world sound scenes despite the presence of reverberation and background noise . GRAM-Ambisonics has the lowest localization error, substantially outperforming the supervised localization model Spatial-AST. For GRAM-Binaural, time-based masking is more successful than patch-based masking. \autoref{fig:local}B shows that GRAMs with spatial attributes (i.e. GRAM-Binaural and -Ambisonics) estimated T60s statistically significantly better than GRAM-Clean and Spatial-AST.

\begin{table}[!t]
\scriptsize
\caption{Performance on real-life recordings of static sound scenes (two tasks in RealSELD). Scores reflect direct transfer without fine-tuning (HEAR pipeline).}
\label{tab:merged_spatial_audio}
\begin{tabular}{l ccc c}
\toprule
& \multicolumn{3}{c}{\textbf{TUT-2018}} & \textbf{TAU-2019} \\ 
 & \textbf{1 source} & \textbf{2 sources} & \textbf{3 sources} & \textbf{2 sources} \\ 
\midrule
\multicolumn{5}{l}{\textit{Detection scores (Error Rate $\downarrow$ / F-score $\uparrow$)}} \\ 
SELDnet & 0.40 / 60.3 & 0.49 / 53.1 & 0.53 / 51.1 & 0.34 / 79.9 \\
MSEDnet & 0.35 / 66.2 & 0.38 / 61.6 & \textbf{0.41} / 59.5 & -- \\
SEDnet & 0.38 / 64.6 & 0.42 / 61.5 & 0.43 / 57.2 & -- \\
\rowcolor[gray]{.95} GRAM-Ambisonics & \textbf{0.30} / \textbf{80.3} & \textbf{0.38} / \textbf{75.0} & 0.43 / \textbf{70.8} & \textbf{0.23} / \textbf{86.4} \\ 
\midrule
\multicolumn{5}{l}{\textit{Localization scores (Localization Error$^\circ \downarrow$ / Frame Recall $\uparrow$)}} \\ 
SELDnet & 26.6 / 64.9 & 33.7 / 41.5 & 36.1 / 24.6 & 28.5 / \textbf{85.4} \\
DOAnet             & \textbf{6.3} / 46.5  & 20.1 / 11.5 & 25.8 / 2.9  & -- \\
Spatial Librispeech   & 12.4 / --   & --          & --          & -- \\
ELSA                      & 15.0 / --   & --          & --          & -- \\  
\rowcolor[gray]{.95} GRAM-Ambisonics    & 11.7 / \textbf{82.1} & \textbf{18.6} / \textbf{49.9} & \textbf{23.0} / \textbf{28.4} & \textbf{12.6} / 79.1 \\
\bottomrule
\end{tabular}
\end{table}

\textbf{SELD task in real-life recordings: } RealSELD contains two tasks with real-life recordings of static scenes: TUT-2018 Real~\citep{seldnet} and TAU-2019~\citep{tau2019}. \autoref{tab:merged_spatial_audio} demonstrates that GRAM-Ambisonics generalizes well to both datasets, achieving lower localization errors than both supervised models trained on in-domain data as well as self-supervised models. Furthermore, \autoref{tab:spatial_audio_regression_2} shows that GRAM-Ambisonics also achieves competitive performance on the datasets comprising real-life recordings of dynamic sound scenes with moving sound sources (TAU-NIGENS-2020 and TAU-NIGENS-2021). GRAM-Ambisonics surpasses the supervised baseline on the TAU-NIGENS-2020 dataset~\citep{tau2020} as presented in the DCASE challenge. Although GRAM-Ambisonics did not outperform the baseline on TAU-NIGENS-2021~\citep{tau2021} (likely due to increased polyphony and the high-intensity directional noise interferers), our results show that even without any fine-tuning or specialized data augmentation during training, GRAM-Ambisonics generalizes well to moving-sound scenes in very challenging real-world conditions.

\begin{figure}[!t]
    \centering
   \includegraphics[width=\linewidth]{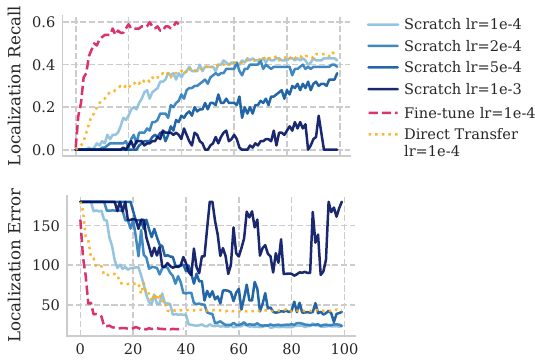}
    \caption{\textbf{Impact of real-life data during training.} Comparing GRAM-Ambisonics validation scores on STARSS23 for training from scratch, fine-tuning a pre-trained model, and direct transfer of the pre-trained model using HEAR pipeline.}
    \label{fig:starss23_training}
\end{figure}

\textbf{Impact of real-life recordings during training: } \autoref{fig:starss23_training} shows that fine-tuning a pre-trained GRAM-Ambisonics model generalizes better and faster to real-life recordings (STARSS23) than training a GRAM-Ambisonics model from scratch on real-world recordings. This highlights that pre-training on simulated naturalistic scenes provides a  strong starting point for efficient generalization to real-life recordings through fine-tuning. 

\begin{figure*}[!htb]
    \centering
    \includegraphics[width=\textwidth]{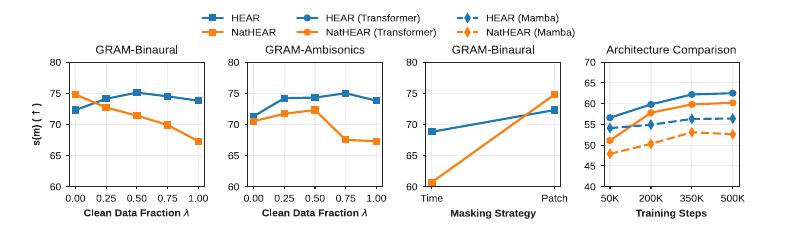}
    \caption{\textbf{Ablation studies}.  Scores reflect downstream task performance across all tasks of the benchmark task suite (y-axis, $s(m)$). From left to right: (1) Impact of mixing clean and naturalistic scenes during pre-training of GRAM-Binaural. (2) Impact of ratio $\lambda$ of clean and naturalistic scenes during pre-training of GRAM-Ambisonics.  (3) Effect of masking strategy for GRAM-Binaural (4) Comparing Mamba and Transformer encoders for binaural audio.}
    \label{fig:merged}
\end{figure*}


\begin{table}[!t]
\scriptsize
\caption{Performance on real-life recordings of dynamic sound scenes (two tasks in RealSELD). Scores reflect direct transfer without fine-tuning (HEAR pipeline).} 
\label{tab:spatial_audio_regression_2}
\begin{tabular}{lcc}
\toprule
& {\textbf{TAU-NIGENS-2020}} & \textbf{TAU-NIGENS-2021} \\ 
\midrule
\multicolumn{3}{l}{\textit{Location-aware detection scores (ER$_{20^{\circ}}$ / F$_{20^{\circ}}$)}} \\ 
Baseline & 0.72 / 37.4 & \textbf{0.73} / \textbf{30.7} \\
\rowcolor[gray]{.95} GRAM-Ambisonics & \textbf{0.57} / \textbf{47.4} & 0.74 / 21.4 \\ 
\midrule
\multicolumn{3}{l}{\textit{Localization scores ($LE_{CD}^\circ \downarrow$ / $LR_{CD} \uparrow$)}} \\ 
Baseline & 22.8 / 60.7 & \textbf{24.5} / \textbf{44.8} \\
\rowcolor[gray]{.95} GRAM-Ambisonics & \textbf{22.3} / \textbf{63.1} & 37.5 / 49.2 \\
\bottomrule
\end{tabular}
\end{table}

\textbf{Mixing clean and naturalistic scenes in pre-training:} We investigated
to what extent pretraining on a mixture of clean and simulated naturalistic sound scenes rather than naturalistic sound scenes only affected the performance of GRAM-Binaural and GRAM-Ambisonics on HEAR and NatHEAR. \autoref{fig:merged} (left panel) shows that performance of GRAM-Binaural on NatHEAR increases with more naturalistic training data (i.e., lower $\lambda$), while performance on HEAR is optimal with a mixture of clean and naturalistic scenes. \autoref{fig:merged} (second panel from the left) shows that GRAM-Ambisonics performed best on both NatHEAR and HEAR with a mixture of clean and naturalistic scenes, consistent with~\citet{elsa}. 

\textbf{Masking strategy:} \autoref{fig:merged} (second panel from the right) illustrates that patch-based masking results in better downstream performance on both HEAR and NatHEAR, although masking strategy had less impact on HEAR performance than NatHEAR performance. 

\textbf{Encoder architecture:} As shown in \autoref{fig:merged}, the Transformer backbone consistently performed better than the Mamba backbone both on clean downstream tasks and on synthetic naturalistic downstream tasks.

\section{Discussion and conclusion}
We present GRAM, a general-purpose, robust spatial audio representation model based on multi-channel MAE. GRAM demonstrates remarkable performance on HEAR and NatHEAR, achieving state-of-the-art results for a self-supervised spectrogram-based audio foundation model while requiring only a fraction of the training data. Moreover, GRAM is the first audio foundation model to perform SELD tasks in real-life recordings. Our experiments demonstrated that GRAM successfully encoded spatial general-purpose audio representations, both in simulated naturalistic scenes and recordings of real-world scenes. Additionally, we release NatHEAR and RealSELD, two new benchmark task suites that provide a standardized manner to evaluate the performance of audio foundation models on simulated naturalistic sound scenes and recordings of real-life sound scenes. In sum, GRAM is a new state-of-the-art audio representation model that incorporates spatial learning and exhibits robust performance in real-world sound scenes, representing a crucial step towards successful applications of audio foundation models in real-world environments.

\textbf{Limitations and future work}: The resolution of mel-spectrograms for binaural inputs was not adequate for learning interaural time differences, which may have hindered the localization performance of the GRAM-Binaural. For future work, GRAM opens the way to multimodal spatial learning. It can serve as a basis for downstream applications such as audio-visual scene representation learning~\citep{clip}, robotics~\citep{ledder2025audio}, and audio-language representation learning \citep{spatialast, chu2023qwen}.

\label{discussion}








\section*{Impact Statement}

This paper presents work whose goal is to advance the field of Machine Learning. There are many potential societal consequences of our work, none which we feel must be specifically highlighted here.

\section*{Acknowledgements}

This project received funding from the NWO Talent Program (VI.Veni.202.184; KH). This work used the Dutch national e-infrastructure with the support of the SURF Cooperative, using grant no. EINF-12218. We would like to thank Robert Jan Schlimbach from the Snellius team for helpful discussions and their help with high-performance cluster utilization. 

\bibliography{example_paper}
\bibliographystyle{icml2026}

\newpage
\appendix
\onecolumn
\section*{Appendix}

\section{Pre-training specification for GRAMs}

We trained all GRAMs for 500K steps on an H100 92GB GPU machine with 16 CPU cores. We used the AdamW optimizer~\citep{loshchilov2017decoupled} with weight decay rate of 0.01, gradient clipping, and a cosine learning rate scheduler with 10 K steps warm-up. The initial learning rate was set to 0.0002, and decayed to 0. We minimize the mean squared error (MSE) between the predicted masked patches and their corresponding input spectrogram patches. GRAM training converges in only 40 GPU hours. 

\section{Evaluated models}
\label{appendix:models}

We compare the performance and efficiency of GRAM-Binaural, GRAM-Ambisonics on downstream tasks with state-of-the-art self-supervised audio representation models with a similar number of parameters as GRAM (90 M): MAE-16x16~\citep{audiomae}, SSAST-patch~\citep{gong2022ssast}, BEATs-iter3~\citep{beats}, MWMAE-B-200-4x16~\citep{mwmae}, SSAM~\citep{audiomamba}; self-supervised speech representation models Wav2Vec 2.0 Base~\citep{wav2vec2}, HuBERT Base~\citep{hubert}, WavLM Base~\citep{wavlm}. To quantify the impact of pre-training with naturalistic sound scenes, we further train GRAM-Clean. GRAM-Clean follows the same experimental setup as GRAM-Binaural and GRAM-Ambisonics, but only uses dry audio clips from the AudioSet. Furthermore, we included Spatial-AST~\citep{spatialast} because it is the only model trained on spatial sound scenes.

\section{Extracting GRAM embeddings for downstream tasks}
\label{gram_embeddings}
We extracted GRAM embeddings for downstream evaluations by encoding embeddings for all patches $P_1, \ldots, P_n$ using the GRAM encoder. We used the exact patch aggregation process as in~\citep{masked-modelling}. Audio clips were split into non-overlapping 2-second chunks and the embedded patches concatenated over time. Later, we took the mean over the time axis to generate scene embeddings independent of the input audio duration. Finally, to evaluate GRAMs on the NatHEAR on localization tasks, we used [CLS] embeddings of the 2-second samples, and averaged them to create scene embeddings for localization tasks.

\section{HEAR and NatHEAR and RealSELD Tasks}
\label{hear_appendix_tasks}
Tables \ref{tab:tasks_overall} and \ref{tab:RealSELD-datasets} illustrate the abbreviations, task descriptions, and the types we have used to benchmark our models.

\begin{table}[!htb]
\renewcommand{\arraystretch}{1.3}
\centering
\caption{Overview of the HEAR and NatHEAR tasks. }\resizebox{\textwidth}{!}{
\begin{tabular}{llll}
\toprule
\textbf{Abbreviation} & \textbf{Task Name} & \textbf{Description} & \textbf{Type}  \\
\midrule
DCASE & DCASE-2016 Task 2~\citep{Mesaros2018_TASLP} &  Event detection of overlapping office sounds in synthetic mixtures & Scene Analysis \\
FS50K & FSD50k~\citep{fsd50k} &  Multilabel, large scale audio tagging & Scene Analysis \\
LC & LibriCount~\citep{libricount} & Speaker Count Identification, Simulated Cocktail Party & Scene Analysis \\
ESC-50 & ESC-50~\citep{esc50} &  Environmental Sound Classification & Environmental Sound Classification \\
CD & Crema-D~\citep{cremad} & Emotion Recognition & Speech Analysis \\
VL & VoxLingua107 Top10~\citep{voxling} & Spoken language identification & Speech Analysis \\
SC-5 & Speech Command 5h~\citep{warden2018speechcommandsdatasetlimitedvocabulary} & Keyword Spotting, reduced training subset & Speech Analysis \\
NS & NSynth Pitch 5h~\citep{nsynth2017} &  Pitch Classification, reduced training subset & Pitch Classification \\
BO & Beijing Opera~\citep{Beijingop} &  Classifying percussion instruments & Percussion \\
Mri-S &  Mridangam Stroke~\citep{mrist} &  Stroke classification in pitched percussion instruments & Percussion \\
Mri-T &  Mridangam Tonic~\citep{mrist} &  Tonic classification in pitched percussion instruments & Percussion \\
\bottomrule
\end{tabular}
}
\label{tab:tasks_overall}
\end{table}

\begin{table*}[!htb]
\centering
\small
\caption{Characteristics of the evaluation datasets for RealSELD.}
\label{tab:RealSELD-datasets}
\begin{tabular}{@{}lllll@{}}
\toprule
\textbf{Name} & \textbf{Motion} & \textbf{Impulse Response} & \textbf{Max Overlap} & \textbf{Noise} \\ \midrule
TUT-2018 Real (ov1, 2, 3) ~\citep{seldnet} & Static & Real & 1, 2, 3 & Ambient 30dB  \\
TAU-2019 ~\citep{tau2019} & Static & Real & 2 & Ambient 30dB \\
TAU-NIGENS-2020 ~\citep{tau2020} & Dynamic & Real & 2 & Ambient [30-6dB] \\
TAU-NIGENS-2021 ~\citep{tau2021} & Dynamic & Real & 3 & Directional interference + Ambient\\
STARSS23 ~\citep{STARSS23} & Dynamic & -- & 6 & Recorded \\ \bottomrule
\end{tabular}
\end{table*}

\section{Downstream performance metric}
\label{appendix:genmetric}

Similar to the procedure in SUPERB \citep{superb}, let $s_t$ be the metric for task $t$. We then calculate the generalizability metric $\text{HEAR}_s(m)$, and $\text{NatHEAR}_s(m)$ for model $m$ as:

\begin{align*}
    s(m) &= \frac{100}{T}\sum^{T}_{t} \frac{s_{t}(m) - s_{t}(baseline)}{s_{t}(SOTA) - s_{t}(baseline)}
\end{align*}

Intuitively, this metric ranks the improvement of models over the baseline as a function of the maximum improvement over the baseline obtained by the current state-of-the-art. Note that we replace $s_{t}(m)$ for task $t$ of model $m$ with 0 when the model scores below baseline performance for task $t$. Similarly, when $s_t(SOTA)$ is lower than baseline for task $t$, we set for all models $s_t$ for this task to 0. In this way, all values are restricted to a range of improvement between 0\% and 100\%.  

\section{Additional Ablation Studies}
\label{appendix:ablations}
Firstly, we further investigated the masking ratio, and in batch sampling as a function of HEAR and NatHEAR performance. Secondly, we investigated the localization performance as a function of the mixture of naturalistic and clean audio ($\lambda$). Thirdly, we investigated the localization performance in terms of noise levels in the NatHEAR-Synethic benchmark, which is low [20-40dB], medium [10-20dB] and high [5-10]dB. Lastly, we examined the effect of in-batch sampling when the effective batch size is held constant. For this experiment, we used gradient accumulation over 16 batches. Consequently, the number of in-batch samples was set to 16, yielding an effective batch size of 512 for both models.

\begin{figure}[!htb]
    \centering
    \includegraphics[width=1\linewidth]{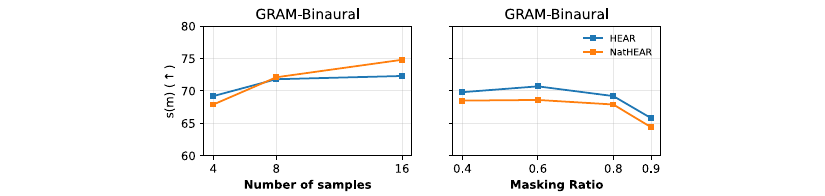}
\caption{Additional ablation studies. Effect of hyperparameters on HEAR and NatHEAR Performance. From
left to right; (1) GRAM-Binaural downstream performance as a function of the number of in batch samples. (2) The effect of masking ratio for GRAM-Binaural. Important to note that GRAM-Binaural depicted in (2) was trained on reduced number of samples (16 $\rightarrow$ 4).}
    \label{fig:appendix-abl}
\end{figure}

\begin{figure}[!htb]
    \centering
    \includegraphics{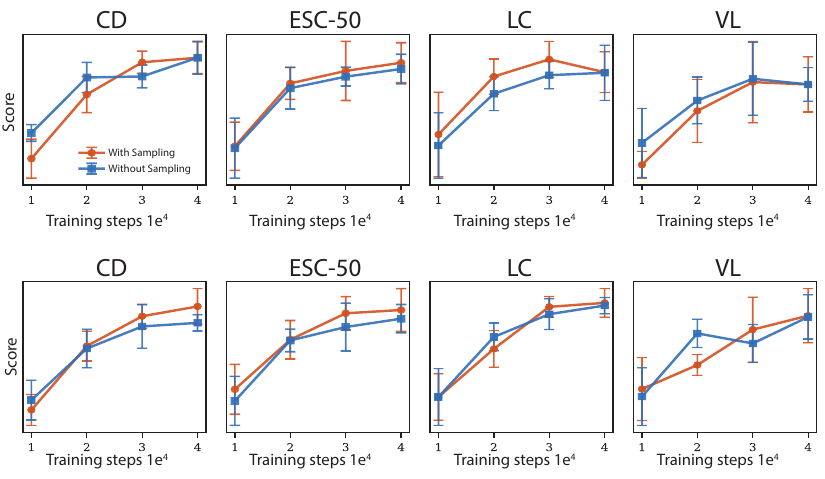}
\caption{Additional ablation studies. Effect of 
in batch sampling on HEAR and NatHEAR performance when the effective batch size is kept the same. From top to bottom; (1) GRAM-Binaural downstream performance on HEAR as a function of the in-batch sampling (2) GRAM-Binaural downstream performance on NatHEAR as a function of the in-batch sampling}
    \label{fig:appendix-abl-hear}
\end{figure}

\textbf{In-batch sampling:} \autoref{fig:appendix-abl} 1 depicts that in-batch sampling helped immensely with the downstream performance on both HEAR and NatHEAR downstream. Increasing the number of in-batch samples leads to higher batch sizes with minimal computational constraints. Furthermore, \autoref{fig:appendix-abl-hear} shows that in-batch sampling does not result in a drop in downstream performance or model convergence.

\textbf{Masking ratio:}  \autoref{fig:appendix-abl} 2 depicts that optimal masking ratio is 0.6 for HEAR and NatHEAR performance, and higher masking ratios, such as 0.9 harms the performance.

\section{Results on original HEAR benchmark suite}

We evaluated our models on the dry, non anechoic, and non-spatialized HEAR Benchmark suite. Table \ref{tab:dry_audio_benchmarks} depicts the achieved results on the HEAR sub tasks.

\begin{table}[!htb]
\renewcommand{\arraystretch}{1.2}
\setlength{\tabcolsep}{3pt}
\caption{Performance comparison of audio representation models across HEAR tasks. All values represent the HEAR scores with standard deviation where available. Bold numbers indicate the best performing model on the specific task. SSAST* is trained on both AudioSet and Librispeech.}
\label{tab:dry_audio_benchmarks}
\centering
\resizebox{\textwidth}{!}{%
\begin{tabular}{l|cccc|ccc|cccc|cc}
\toprule
& \multicolumn{4}{c|}{\textbf{Acoustic Events and Scene Analysis}}  & \multicolumn{3}{c|}{\textbf{Speech}} & \multicolumn{4}{c|}{\textbf{Music}} & \multicolumn{1}{c}{}\\
\textbf{Model} & \textbf{DCASE} & \textbf{FSD50K} & \textbf{LC} &  \textbf{ESC-50} & \textbf{CD} & \textbf{VL} & \textbf{SC-5}   & \textbf{NS} &  \textbf{BO} & \textbf{Mri-S} & \textbf{Mri-T} & \textbf{s(m)} & \textbf{Avg.}\\
\midrule
\multicolumn{12}{l}{\textbf{Baseline}} \\
HEAR-Naive & 8.8 & 13.2 & $ 43.5 \pm 1.6 $  & $ 28.6 \pm 3.1 $  & $ 38.0 \pm 2.3 $  & $ 14.8 \pm 3.0 $  & 13.3 & 87.6 & $ \textbf{98.7} \pm 1.9 $ & $ 94.1 \pm 0.5 $  & $ 87.6 \pm 6.4 $  & 0.0 & 48.0\\
\midrule
\multicolumn{13}{l}{\textbf{Speech SSL}} \\
Wav2Vec 2.0 & 23.5 & 29.4 & $ 69.9 \pm 2.1 $  & $ 46.4 \pm 1.8 $  & $ 57.3 \pm 1.1 $  & $ 34.9 \pm 2.4 $  & 85.3 & 17.4 & $ 81.4 \pm 4.8 $  & $ 90.7 \pm 0.8 $  & $ 77.0 \pm 0.9 $ & 30.7 & 55.7\\
HuBERT & 78.3 & 32.8 & $ 63.3 \pm 1.2 $  & $ 58.6 \pm 2.8 $  & $ 71.2 \pm 1.2 $  & $ \textbf{65.2} \pm 2.9 $  & \textbf{94.0} & 19.8 & $ 93.2 \pm 5.9 $  & $ 94.6 \pm 0.4 $  & $ 85.0 \pm 2.5 $ & 43.6 & 68.7  \\
WavLM & 27.0 & 25.7 & $ 61.3 \pm 2.3 $  & $ 49.5 \pm 3.8 $  & $ 64.3 \pm 1.3 $  & $ 60.1 \pm 3.2 $  & 93.8 & 18.2 & $ 84.3 \pm 6.3 $  & $ 88.8 \pm 1.0 $  & $ 76.8 \pm 0.5 $ & 36.1 & 59.1\\
\hline
\multicolumn{13}{l}{\textbf{AudioSet SSL}} \\
MAE & --  & 33.4 & $ 62.3 \pm 1.1 $  & $ 72.9 \pm 2.1 $  & $ 60.8 \pm 1.8 $  & $ 21.3 \pm 5.8 $  & 66.6 & 63.6 & $ 94.5 \pm 5.6 $  & $ 94.8 \pm 0.6 $  & $ 85.1 \pm 10.4 $  & 31.3 & 65.5\\
SSAST* & --  & 21.4 & $ 57.8 \pm 3.3 $  & $ 58.3 \pm 2.6 $  & $ 48.0 \pm 2.1 $  & $ 15.4 \pm 2.6 $  & 22.0 & 64.2 & $ 95.8 \pm 4.3 $  & $ 90.2 \pm 5.9 $  & $ 89.1 \pm 8.0 $ & 15 & 56.2 \\
BEATs & --  & 54.1 & $ 77.8 \pm 1.2 $  & $ 85.8 \pm 2.9 $  & $ 66.9 \pm 2.5 $  & $ 39.7 \pm 4.3 $  & 86.9 & 68.6 & $ 94.1 \pm 3.5 $  & $ 95.5 \pm 0.4 $  & $ 96.6 \pm 0.5 $ & 59.2 & 76.6 \\
MWMAE & 94.2 & 51.8 & $ 80.3 \pm 1.9 $  & $ 82.2 \pm 3.2 $  & $ 74.4 \pm 1.5 $  & $ 45.5 \pm 1.7 $  & 91.6 & 69.4 & $ 95.8 \pm 4.3 $  & $ 97.5 \pm 0.4 $  & $ 97.6 \pm 0.6 $ & 68.9 & 80.8\\
SSAM & 87.3 & 53.5 & $ 75.5 \pm 1.4 $  & $ 82.9 \pm 3.6 $  & $ 70.2 \pm 0.4 $  & $ 56.4 \pm 5.2 $  & 89.3 & 72.6 & $ 93.2 \pm 3.5 $  & $ \textbf{97.8} \pm 0.5 $  & $ 96.9 \pm 0.5 $ & 69.0 & 79.6  \\
GRAM-Binaural & \textbf{95.6} & 56.1 & $81.0 \pm 1.1 $  & $ 86.7 \pm 2.4 $  & $ 75.0 \pm 1.4 $ & $ 53.2 \pm 3.0 $  & 92.5 & \textbf{77.0} & $ 94.9 \pm 3.2 $  & $ 97.3 \pm 0.3 $  & $ 98.1 \pm 0.2 $ & 72.3 & 82.5 \\
GRAM-Ambisonics & 94.3 & 53.0 & $ 79.4 \pm 1.5 $ & $ 85.9 \pm 1.5 $ & $ 71.9 \pm 1.9 $ & $ 53.7 \pm 1.2 $ & 89.6 & 73.8 & $ 94.9 \pm 4.9 $ & $ 97.6 \pm 0.5 $ & $\textbf{98.5} \pm 0.4 $ & 71.3 & 81.1\\
GRAM-Clean & 95.3 & 56.8 & $\textbf{81.3} \pm 1.8 $ & $ \textbf{87.5} \pm 2.3 $ & $ \textbf{75.1} \pm 0.6 $ & $ 57.3 \pm 3.4 $ & 93.5 & 75.8 & $ 95.8 \pm 3.7 $ & $ 97.4 \pm 0.3 $ & $ 98.0 \pm 0.2 $& \textbf{73.8} & \textbf{83.1}\\
\midrule
{\textbf{Supervised}} \\
PASST & --  & \textbf{64.1} & $ 60.7 \pm 3.7 $  & $ \textbf{94.8} \pm 0.3 $  & $ 61.8 \pm 1.1 $  & $ 25.9 \pm 2.6 $  & 68.7 & 24.2 & $ \textbf{96.6} \pm 3.2 $  & $ 96.4 \pm 0.7 $  & $ 87.8 \pm 1.2 $ & 46.2 & 68.1\\
Spatial-AST & --  & 54.7 & $ 72.6 \pm 1.5 $  & $ 90.3 \pm 1.7 $  & $ 62.2 \pm 1.3 $  & $ 29.1 \pm 1.9 $  & 80.6 & 69.8 & $ 96.2 \pm 5.3 $  & $ 96.2 \pm 0.4 $  & $ 94.6 \pm 0.6 $ & 54.6 & 74.6 \\
\bottomrule
\end{tabular}
}
\end{table}

\section{SELD Training with ACCDOA}

We extend the HEAR Benchmark with ACCDOA ~\citep{shimada2021accdoa} framework to solve newly introduced real world SELD tasks. This framework jointly models sound event detection and localization across target sound classes. A class is considered active at frame $t$ when the predicted Cartesian coordinate magnitude $\|\mathbf{c}_t\| > 0.5$, where $\mathbf{c}_t \in \mathbb{R}^3$ represents the unit direction vector. We do not perform any post-processing steps unlike the HEAR protocol.

To adapt the datasets to the HEAR format, we processed the ground truth timestamp labels into fixed-duration segments. For static sources, we extracted start and end times alongside fixed azimuth and elevation coordinates. For moving sources, we extracted active segments and their corresponding time-varying directions. In frames where no sources are active, the ACCDOA target is defined as a zero vector. Crucially, the dynamic datasets in RealSELD utilize 100ms segments (10 Hz resolution). To align the output resolution of the audio embedding model with these labels, we applied average pooling to the model representations. This ensures the GRAM model output is temporally aligned with the 100ms ground truth segments.

After the mapping, we extract the time-stamp embeddings and their corresponding ground truth labels using the HEAR evaluation kit, and linear probe the extracted information using the HEAR evaluation protocol.

\label{appendix:accdoa}

\section{Evaluation Metrics}

We evaluate the sound localization performance on the newly generated sound localization tasks in NatHEAR by calculating the Direction of Arrival (DoA) error $\theta$ between the [$x,y,z$] coordinates of the target sound source on the unit sphere using the arc cosine of the dot product of the unit vectors: $\theta = \arccos(v \cdot  \hat{v})$.

\textbf{SELD metrics on RealSELD Dynamic Motion: } We used four joint localization and detection metrics ~\citep{eval_overview}, which are widely-used in audio-only SELDtasks ~\citep{STARSS22}. Two metrics, referred to as location-aware detection, are the error rate (ER$_{20^{\circ}}$) and F-score (F$_{20^{\circ}}$) for one-second non-overlapping segments. We consider a prediction to be a true positive (TP) if the prediction and the reference class are the same and the angle difference is less than $20^{\circ}$. F$_{20^{\circ}}$ is calculated from location-aware precision and recall, whereas ER$_{20^{\circ}}$ is the sum of insertion, deletion, and substitution errors, divided by the total number of references. The other two metrics, referred to as class-aware localization, are localization error (LE$_{CD}$) in degrees and localization recall (LR$_{CD}$) in one-second non-overlapping segments, where the subscript denotes classification-dependent. Unlike location-aware detection, we do not use a threshold; instead, we estimate the difference between the correct prediction and the reference. LE$_{CD}$ expresses the average angular difference between the same class’s predictions and references. LR$_{CD}$ reports the true positive rate as the number of localization estimates detected in a class out of the total number of class instances. We used the macro mode of computation, which does not apply to ER$_{20^{\circ}}$ because it includes substitution errors between two classes. We first computed the metrics for each class and then averaged them for the other three metrics to obtain the final system performance. 

\textbf{SELD metrics on RealSELD Static Motion: }For testing the model on the TUT-2018 REAL and TAU-2019, we used the non-localization-dependent counterparts of the aforementioned metrics. ER and F-score calculated in segments of one second with no overlap, as proposed. The segment-wise results are obtained from the classifier's frame-level predictions, treating a sound event as active across the entire segment if it is active in any frame. Similarly, we obtain labels for one-second segments of the reference from its framewise annotation and calculate segment-wise ER and F-scores. Additionally, in order to account for time frames where the number of estimated and reference DOAs are unequal, we report the frame recall, calculated as TP (TP + FN) in percentage, where true positives TP is the total number of time frames in which the number of DOAs predicted is equal to reference, and false negatives FN is the total number of frames where the predicted and reference DOA are unequal.

\section{T60 estimation tasks}

NatHEAR includes two T60 estimation tasks (ESC-50 and SC-5) in addition to the direction-of-arrival estimation tasks. For synthesizing these tasks, we did not add additional localized/diffused noise. Specifically, we convolved ESC-50 and SC-5 clips with BRIR$(s,r,\theta )$ for NatHEAR Binaural, or with ARIR$(s,r,\theta )$ for NatHEAR Ambisonics. We estimated the ground truth RT60s using the first channel of the ARIR. To estimate the T60s, we utilized the Schroeder method ~\citep{rir-method} from Pyroomacoustics package ~\citep{pyroomacoustics}. Explicitly, we measure the RT30 and extrapolate to T60 using the decay curve.

\begin{figure}[!htb]
    \centering
    \includegraphics{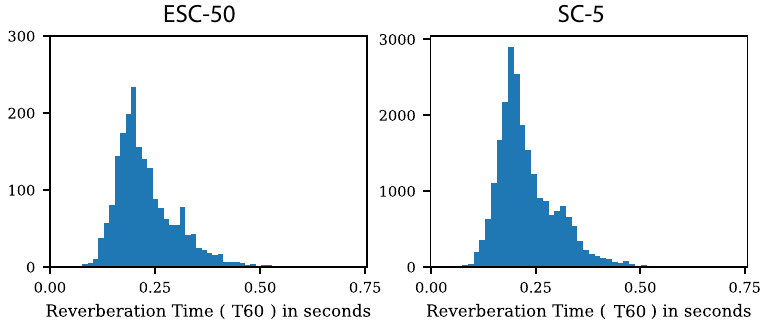}
    \caption{Distribution of the estimated T60s for ESC-50 and SC-5 datasets.}
    \label{appendix:rt60s}
\end{figure}

~\autoref{appendix:rt60s} depicts the RT60 distributions of ESC-50 and SC-5 datasets. Furthermore ~\autoref{tab:errors} presents the median absolue errors that we got with GRAMs on ESC-50 and SC-5 tasks.

\begin{table}[h]
\centering
\caption{Absolute median error comparison on T60 estimation tasks. }
\begin{subtable}[t]{0.45\textwidth}
\centering
\caption{SC-5}
\begin{tabular}{lc}
\toprule
\textbf{Model} & \textbf{Median Error} \\
\midrule
GRAM-T-Clean & 0.0225 \\
GRAM-T-Ambisonics & 0.0169 \\
GRAM-T-Binaural (Patch) & 0.0146 \\
GRAM-T-Binaural (Time) & 0.0179 \\
Spatial-AST & 0.0299 \\
\bottomrule
\end{tabular}
\label{tab:errors1}
\end{subtable}
\hfill
\begin{subtable}[t]{0.45\textwidth}
\centering
\caption{ESC-50}
\begin{tabular}{lc}
\toprule
\textbf{Model} & \textbf{Median Error} \\
\midrule
GRAM-T-Clean & 0.0461 \\
GRAM-T-Ambisonics & 0.0421 \\
GRAM-T-Binaural (Patch) & 0.0397 \\
GRAM-T-Binaural (Time) & 0.0418 \\
Spatial-AST & 0.0468 \\
\bottomrule
\end{tabular}
\label{tab:errors2}
\end{subtable}
\label{tab:errors}
\end{table}

\section{Evaluating training efficiency}

For all models trained solely on AudioSet, we calculated the number of seconds seen during the training as: batch size $\times$ steps per epoch $\times$ epochs $\times$ input length. This comparison accounts for the number of 10-second AudioSet sound clips processed by each model.

\begin{table}[!htb]
\scriptsize
\centering
\caption{Training details of the recent audio foundation models. We retrieve the numbers from the references where possible. Various works utilized various sizes of AudioSet. Therefore, we used the dataset size reported by the references to calculate the steps per epoch. For MWMAE and SSAM we retrieved their dataset size from their corresponding code repository.}\resizebox{\textwidth}{!}{
\begin{tabular}{lccccc}
\toprule
\textbf{Model} & \textbf{Batch Size} & \textbf{Epochs} & \textbf{Steps per Epoch} & \textbf{Total Steps} & \textbf{Total Samples Seen} \\
\midrule
MWMAE~\citep{mwmae} & 1024 & 100 & 1985 & 198500 & $\sim\text{200M}$ \\
GRAMs & 96 & N/A & N/A & 500000 & $\sim\text{48M}$ \\
Audio-MAE~\citep{audiomae} & 512 & 32 & 3829 & 122528 & $\sim\text{48M}$ \\
BEATs~\citep{beats}& 5600 & N/A & N/A & 1.2M & $\sim\text{6.7B}$\\
SSAM~\citep{audiomamba} & 1024 & 100 & 2003 & 200300 & $\sim\text{207M}$ \\

\bottomrule
\end{tabular}
}
\label{tab:training_details}
\end{table}

\pagebreak
\section{SoundSpaces 2.0 specifications}
\label{appendix: brir-spec}

We generate our BRIRs using the simulator provided by SoundSpaces 2.0~\citep{chen22soundspaces2}. Our hyperparameters for the simulator is depicted in Table \ref{tab:acoustic_config}.

\begin{table}[!htb]
\caption{Acoustic configuration parameters utilized in SoundSpaces 2.0 to generate our BRIRs.}
\label{tab:acoustic_config}
\centering
\small
\begin{tabular}{lcllc}
\toprule
\textbf{Parameter} & \textbf{Value} & \phantom{xxxx} & \textbf{Parameter} & \textbf{Value} \\
\midrule
directSHOrder & 3 && indirectSHOrder & 3 \\
sampleRate & 32000 && frequencyBands & 8 \\
maxDiffractionOrder & 10 && transmission & True \\
indirect & True && indirectRayCount & 15000 \\
indirectRayDepth & 400 && sourceRayCount & 200 \\
sourceRayDepth & 20 && threadCount & 16 \\
agentHeigth & 1.5m \\
\bottomrule
\end{tabular}
\end{table}

\end{document}